\documentclass{sigchi}

\CopyrightYear{2019}
\setcopyright{acmlicensed}
\doi{https://doi.org/10.1145/3311350.3347183}
\isbn{978-1-4503-6688-5/19/10}
\conferenceinfo{CHI PLAY '19,}{October  22--25, 2019, Barcelona, Spain}
\acmPrice{\$15.00}

\usepackage{balance}       %
\usepackage{graphics}      %
\usepackage[T1]{fontenc}   %
\usepackage{txfonts}
\usepackage{mathptmx}
\usepackage[pdflang={en-US},pdftex]{hyperref}
\usepackage{color}
\usepackage{booktabs}
\usepackage{textcomp}

\usepackage{microtype}        %
\usepackage{ccicons}          %
\newcommand{\comm}[1]{}

\usepackage{todonotes}

\def\plaintitle{Outstanding: A Multi-Perspective Travel Approach\\ for Virtual Reality Games}

\def\emptyauthor{}
\def\plainkeywords{Virtual reality games; navigation; perspectives; virtual avatar; orientation; virtual body size; world-in-miniature}

\makeatletter
\def\url@leostyle{%
  \@ifundefined{selectfont}{
    \def\UrlFont{\sf}
  }{
    \def\UrlFont{\small\bf\ttfamily}
  }}
\makeatother
\def\pprw{8.5in}
\def\pprh{11in}

\definecolor{linkColor}{RGB}{6,125,233}
\hypersetup{%
  pdftitle={\plaintitle},
  pdfauthor={\emptyauthor},
  pdfkeywords={\plainkeywords},
  pdfdisplaydoctitle=true, %
  bookmarksnumbered,
  pdfstartview={FitH},
  colorlinks,
  citecolor=black,
  filecolor=black,
  linkcolor=black,
  urlcolor=linkColor,
  breaklinks=true,
  hypertexnames=false
}

\begin{document}

\title{\plaintitle}
\comm{
\numberofauthors{3}
\author{%
  \alignauthor{d\\
    \affaddr{c}\\
    \affaddr{b}\\
    \email{a}}\\
\alignauthor{d\\
    \affaddr{c}\\
    \affaddr{b}\\
    \email{a}}\\
\alignauthor{d\\
    \affaddr{c}\\
    \affaddr{b}\\
    \email{a}}\\
}}
\numberofauthors{1}
\author{%
  \alignauthor{Sebastian Cmentowski, Andrey Krekhov, Jens Kr\"uger\\
    \affaddr{High Performance Computing Group}\\
    \affaddr{University of Duisburg-Essen, Germany}\\
    \email{\{sebastian.cmentowski, andrey.krekhov, jens.krueger\}@uni-due.de}}\\
}

\teaser{
  \includegraphics[width=\linewidth]{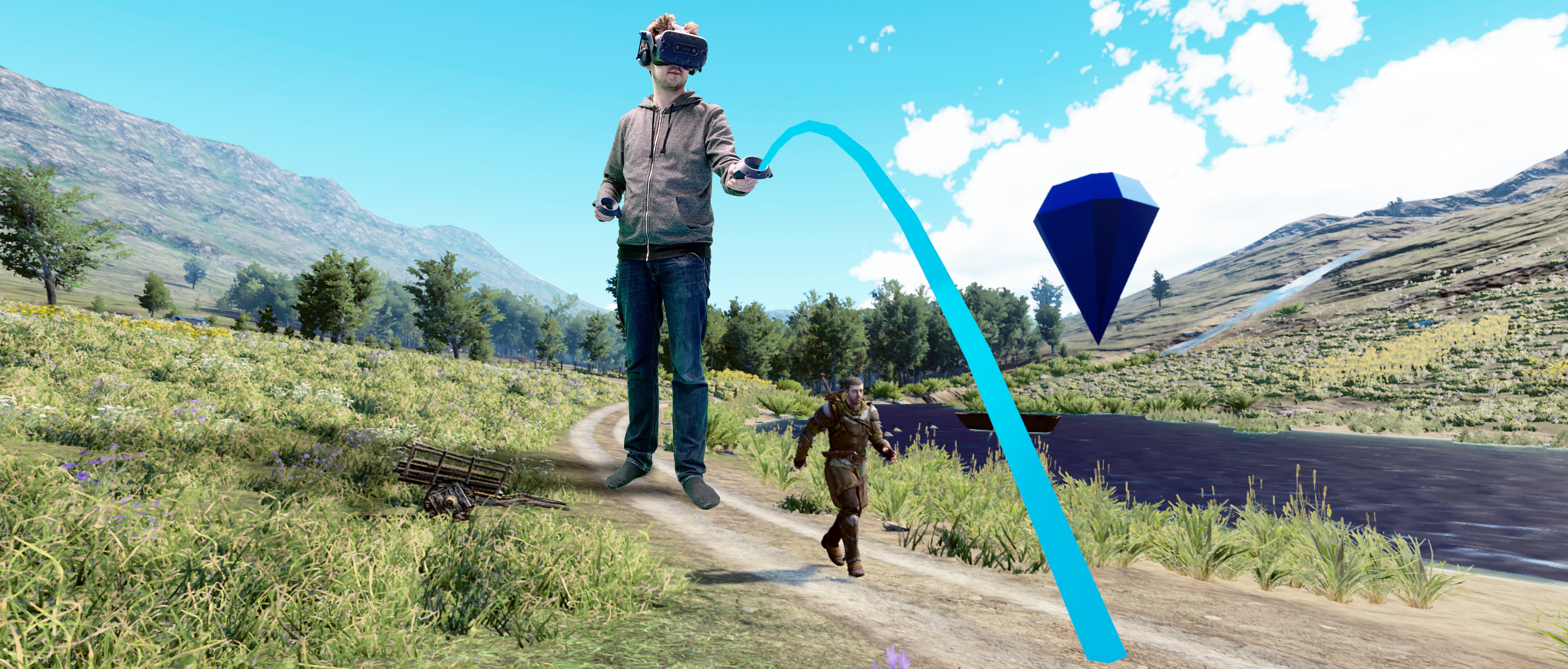}
  
  \caption{Our proposed navigation technique allows players to switch to a scaled third-person perspective on demand and control a virtual avatar to cover large distances in open world VR scenarios.}
  \label{fig:teaser}
}
\maketitle

\begin{abstract}
In virtual reality games, players dive into fictional environments and can experience a compelling and immersive world. State-of-the-art VR systems allow for natural and intuitive navigation through physical walking. However, the tracking space is still limited, and viable alternatives \comm{or extensions }are required to reach further virtual destinations. Our work focuses on the exploration of vast open worlds -- an area where existing local navigation approaches such as the arc-based teleport are not ideally suited and world-in-miniature techniques potentially reduce presence. We present a novel alternative for open environments: Our idea is to equip players with the ability to switch from first-person to a third-person bird's eye perspective on demand. From above, players can command their avatar and initiate travels over large distance. Our evaluation reveals a significant increase in spatial orientation while avoiding cybersickness and preserving presence, enjoyment, and competence. We summarize our findings in a set of comprehensive design guidelines to help developers integrate our technique.
\end{abstract}




\section{Introduction}
Virtual reality allows players to explore fictional environments in an immersive and natural manner, to experience a feeling of being there, and almost to forget the real surrounding. Continuous technical improvements and faster rendering approaches make it possible to push the boundaries of VR even further and develop vast open environments that could be explored freely and immersively. However, large and detailed VR worlds require proper techniques to travel these landscapes.\par
\begin{figure}
\centering
\includegraphics[width=1.0\columnwidth]{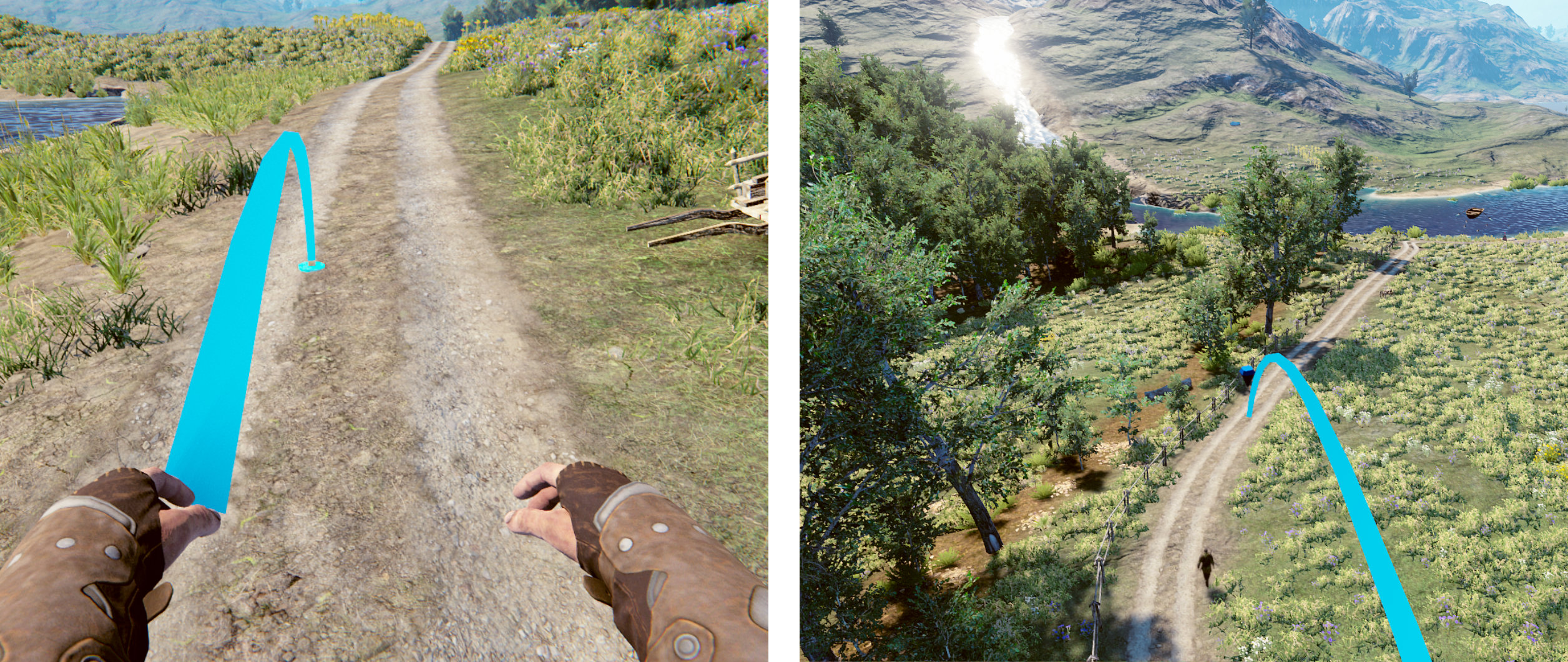}
\caption{Player's perspective in the two study conditions: arc-based teleport (left) and perspective switching (right).}
\label{fig:teleport}
\end{figure}

Physical walking using room-scale tracking offers an intuitive and natural kind of navigation~\cite{ruddle2009benefits}. However, the available walking space is usually confined to the size of a living room. Game developers overcome this limitation by adding virtual locomotion techniques such as the prominent teleport (see Figure~\ref{fig:teleport}). Most of these approaches were designed for local navigation and are not ideally suited for exploring large and open worlds. Only a few exceptions exist, such as the world-in-miniature (WIM)~\cite{stoakley1995virtual}, where players use a miniature model of the virtual scenario to teleport themselves to distant places. Nevertheless, this approach relies on an artificial user interface and does not provide an opportunity to explore an environment freely and continuously which potentially reduces the players' possibility to immerse themselves in the virtual world.\par
Our research closes the gap between local teleportation and WIM relocation by introducing a novel approach for continuous long-distance traveling. Our main idea is to switch dynamically between a first-person and a third-person bird's eye perspective on demand. The first-person mode offers a familiar experience and is used to explore the local surrounding and interact with the environment. In the third-person mode, players see and command their avatar from a bird's eye perspective, as depicted in Figure~\ref{fig:teaser}.\par
Using the correct perspective for every situation offers important benefits~\cite{gorisse2017first}: first-person is suited best for interaction-intensive tasks while third-person provides a better overview. We combine both perspectives to achieve an intuitive navigation approach. Moreover, we extend this basic concept by additional features to enhance the experience further: Virtual scaling of the player in third-person mode is used to improve the spatial orientation and deliver a feeling of moving through a miniature world while commanding an avatar. Additionally, we use a smooth and fast transformation between both perspectives to prevent cybersickness and to emphasize the impression of leaving and re-embodying the virtual avatar.\par
Our main contribution is the proposed navigation technique using dynamic perspective switching. We validate this approach by comparing it against the arc-based teleport using a 3D adventure game. Our experiments reveal significant benefits to spatial orientation and overview while preserving equal levels of presence, enjoyment, and competence. Additionally, our improvements prevent adverse effects through cybersickness. As the final step, we discuss the unique strengths and weaknesses of our proposed approach and condense these into a set of design implications that can help developers and practitioners.\comm{ to use our insights for their purposes.}

\section{Related Work}
Our work belongs to the virtual reality research with a particular focus on VR games and locomotion techniques. Consequently, we first introduce basic concepts and issues behind VR games such as immersion, presence, and cybersickness. Subsequently, we outline the current state of the art in VR locomotion research. Since our technique centers around the concept of perspective-switching and dynamic virtual rescaling, we also provide the necessary background to these topics.\par
Two concepts seem to be of utmost importance when dealing with VR applications: \textit{immersion}~\cite{cairns2014immersion} and \textit{presence}~\cite{heeter1992being}. To stay in line with the majority of recent research, we use the term immersion to describe the technical quality of a VR setup~\cite{Biocca:1995:IVR:207922.207926, sherman2002understanding}. Immersive setups can induce a feeling of being there, which is commonly called presence. This distinction is further formalized by Slater et al.~\cite{slater2003note}, Lombard et al.~\cite{lombard1997heart} and IJsselsteijn et al.~\cite{IJsselsteijn}. For a particular focus on locomotion-related presence, we point to the work by Slater et al.~\cite{slater1995taking}.

\subsection{Cybersickness}
A typical problem most VR applications have to tackle is the occurrence of cybersickness~\cite{laviola2000discussion}. Even though often being used synonymously with the effect of simulator sickness~\cite{kolasinski1995simulator}, both are different strains of the motion sickness phenomenon~\cite{money1970motion,hettinger1992visually,ohyama2007autonomic}. Typical symptoms such as headaches, eye strain, sweating, nausea or vomiting arise due to a mismatch of our vestibular-ocular system.\par
Humans sense acceleration using their vestibular system which usually matches the sensory input gathered from the visual system~\cite{laviola2000discussion}. In the case of a mismatch between these signals, the resulting symptoms differ in strength and form~\cite{reason1975motion}. The reason for this body reaction remains unsolved, but so far three major prominent theories have been established: sensory conflict theory (most accepted), poison theory, and postural instability theory~\cite{laviola2000discussion}. The difference between both specific strains of motion sickness was extensively explored by Stanney et al.~\cite{stanney1997cybersickness}: Simulator sickness usually occurs when a simulator, typically used for pilot or astronaut training, is not correctly configured ~\cite{kennedy1989simulator}. This technical problem can lead to rather mild oculomotor and nausea symptoms. In contrast, cybersickness is caused by a broad set of reasons ranging from technological issues such as flickering and lags to a wrong visual image being caused by mismatches in movement, eye distance or vergence. The results are mainly severe symptoms such as disorientation and nausea~\cite{stanney1997cybersickness}.\par
Additionally, Hettinger et al.~\cite{hettingerVection} introduced the phenomenon of vection as a possible source for cybersickness. Vection is a feeling of moving that is solely induced by the visual system and usually experienced when sitting on a standing train and watching the adjacent train accelerating. This effect is \comm{caused or }supported by different factors listed by La Viola Jr~\cite{laviola2000discussion}: the field of view (FOV) of the HMD, the optical flow rate, the degree of movement and proximity of objects. In short, close and fast-moving objects filling the player's view combined with a big field of view tend to amplify the amount of perceived vection and potential cybersickness. Consequently, Fernandes et al.~\cite{fernandes2016combating} propose limiting the FOV to reduce cybersickness. A broader discussion about the influence of the FOV on cybersickness can be found in the work of Lin et al.~\cite{lin2002effects}.\par
Additionally, recent studies have shown that the accumulated flow over time, perceived via central and peripheral vision, forms a critical factor in the occurrence of motion sickness~\cite{Lee:2017:ESS:3145690.3145697}. Instead of avoiding cybersickness at all costs, von Mammen et al.~\cite{von2016cyber} showed that games with artificially induced cybersickness can still be enjoyable. This leads to the conclusion that a reduction of potential motion sickness to an acceptable level could be more favorable than limiting the opportunities of virtual reality to avoid risking any symptoms.

\subsection{Locomotion}
Most non-VR games use joysticks to control the player's avatar. Such approaches involving continuous motion are rarely transferable to VR as they tend to induce cybersickness~\cite{Habgood:2017:HLP:3130859.3131437}. Alternatively, VR games can use natural walking~\cite{ruddle2009benefits} to achieve intuitive and presence-preserving navigation. However, the confined space of currently available room-scale tracking limits natural walking to a few square meters. \par
Recent research focused on overcoming this limitation by extending the range of real walking to enable the player to reach further. Bhandari et al.~\cite{ Bhandari:2017:LSW:3139131.3139133} combined walking with walking in place~\cite{slater1995taking, tregillus2016vr} and reported higher presence compared to traditional controller input. This result is in line with the work by Usoh et al.~\cite{usoh1999walking}: According to their research, walking is superior to walking in place, while both outperform virtual locomotion\comm{ techniques}. Another approach by Bolte et al.~\cite{bolte2011jumper} uses the detection of physical jumping: When \comm{an acceleration and }a jump is detected, the \comm{resulting }forward motion is augmented to travel larger distances.\par
In contrast to these augmented walking approaches, purely virtual navigation techniques sacrifice the advantages of natural walking to achieve unlimited traveling. A typical problem that arises from the necessary decoupling of real and virtual movement is an increase in cybersickness. This is best tackled by "short, fast movements in VR (with no acceleration or deceleration)"~\cite{Habgood:2017:HLP:3130859.3131437}, which has been confirmed by the work of Medeiros et al.~\cite{medeiros2016effects} and Yao et al.~\cite{yao2014oculus}. The most prominently used navigation approach, using such short movements, is the arc-based teleportation technique: players aim at an accessible destination and are directly teleported there. This approach is superior to the traditional gamepad locomotion~\cite{frommel2017effects} and is actively promoted and encouraged by the majority of established VR systems such as the HTC Vive~\cite{vive}. However, the perceived presence and spatial orientation are significantly lowered by instant relocations. Even worse, the necessity to see the target location limits the maximal distance to be traversed in one jump and vastly increases the necessary workload for more considerable travels or occluded areas. \par
Apart from virtual travel techniques, a couple of other solutions for infinite locomotion have been developed. One famous approach is the extension of available walking space by unconsciously altering the virtual movement from the real walk. Users are not able to sense slight rotations in the virtual environment leading to a feeling of walking on a straight line, whereas in reality, they are moving in circles. However, \comm{ the subliminal turning rate must be minimal which leads to a large circle diameter. The required space exceeds living room sizes by an order of magnitude and would only fit into locations like storage halls.} the necessary minimal turning rate leads to extensive space requirements. This impediment is the main reason why the concept of \textit{redirected walking}~\cite{razzaque2005redirected,razzaque2001redirected} has stayed a pure research topic despite numerous improvements~\cite{engel2008psychophysically,Grechkin:2016:RDT:2931002.2931018,langbehn2016subliminal}.\par
One remedy for true large distance travel in VR is the concept of a world-in-miniature (WIM)~\cite{stoakley1995virtual}: a virtual three-dimensional minimap is shown on players' hands and can be used to move instantaneously to any point within a large and complex environment. This concept has been further refined by La Viola Jr et al.~\cite{laviola2001hands} to achieve a walkable minimap that is grown around the player's feet to replace the previous environment. In comparison to teleport, WIM works better with larger distances and occlusions~\cite{berger2018wim}. However, it introduces the minimap as an additional artificial interface which is decoupled from the original virtual world. Since both approaches were never designed to be a perfect solution for long-distance travel, this encouraged us to develop a possible alternative.\par
\subsection{Perspective and Scale}
Our approach is based on switching between first-person (1PP) and third-person (3PP) perspectives. After early studies on the potential use of 3PP in virtual environments~\cite{salamin2006benefits}, Gorisse et al.~\cite{gorisse2017first} administered the perceptual differences in an extensive study: According to their experiments, both perspectives are able to preserve high levels of presence and agency. However, 1PP is best suited for interaction-intensive tasks\comm{ and scenarios involving body ownership. In contrast,} while 3PP provides advantages to spatial awareness and environmental perception. These findings support our idea for a navigation metaphor using 3PP for large-distance navigation and 1PP for local interaction. Gorisse et al.~\cite{gorisse2017first} decided to place the 3PP viewpoint directly behind the avatar. However, this is not suitable in our case as it does not improve the view distance or environmental knowledge of the user. Instead, we have decided to scale the disembodied players to giant size, similar to the work by Abtahi et al.~\cite{abtahi2019m}. The virtual camera position is moved to a greater height and enables players to perceive the environment as a miniature world. Meanwhile, their avatar resides at his original size to the feet of the players.\par
Dynamic scaling of the virtual world or the player is not new and mostly used within so-called multiscale virtual environments (MSVE)~\cite{Zhang:2002:SIM:571878.571884}. Kopper et al.~\cite{kopper2006design} used MSVEs to explore the inner organs of virtual human bodies and reported that automatic scaling outperforms a manual-chosen scaling factor regarding usability. Similarly, Argelaguet et al.~\cite{argelaguet2016giant} emphasized the importance of automatic scaling speeds and optimized stereoscopic rendering parameters to minimize adverse side effects such as diplopia or cybersickness. While MSVEs use different scaling factors to access distinct observation levels within their virtual environment, our technique focuses on the locomotion aspect and only resides on the scaling as a means to improve visibility and overview. In this manner, the closest resembling approach is the GulliVR technique by Krekhov et al.~\cite{krekhov2018gullivr} that focused on full-body scaling in the first-person perspective. One widely proposed request from the CHIplay-community was to combine this approach with perspective-switching to overcome the limited field of use and to decouple avatar and player. This wish was a major motivation in developing our presented technique.\par
A core aspect of our idea is a bird's eye view on a miniature world. Following previous work, we decided to scale the stereo camera separation accordingly to the rest of the body. While smaller variations of this virtual eye distance do not have any measurable impact on size judgments~\cite{best1996perceptual}, larger differences lead to a \textit{false eye separation}~\cite{cho2014evaluating}. The result is an altered size perception that produces the desired miniature world effect. Additionally, the resulting impression closely matches the perceived virtual movement that the scaled players experience and has been shown to avoid inducing cybersickness~\cite{krekhov2018gullivr}. 

\section{Navigation Technique}
\begin{figure}
\centering
\includegraphics[width=1.0\columnwidth]{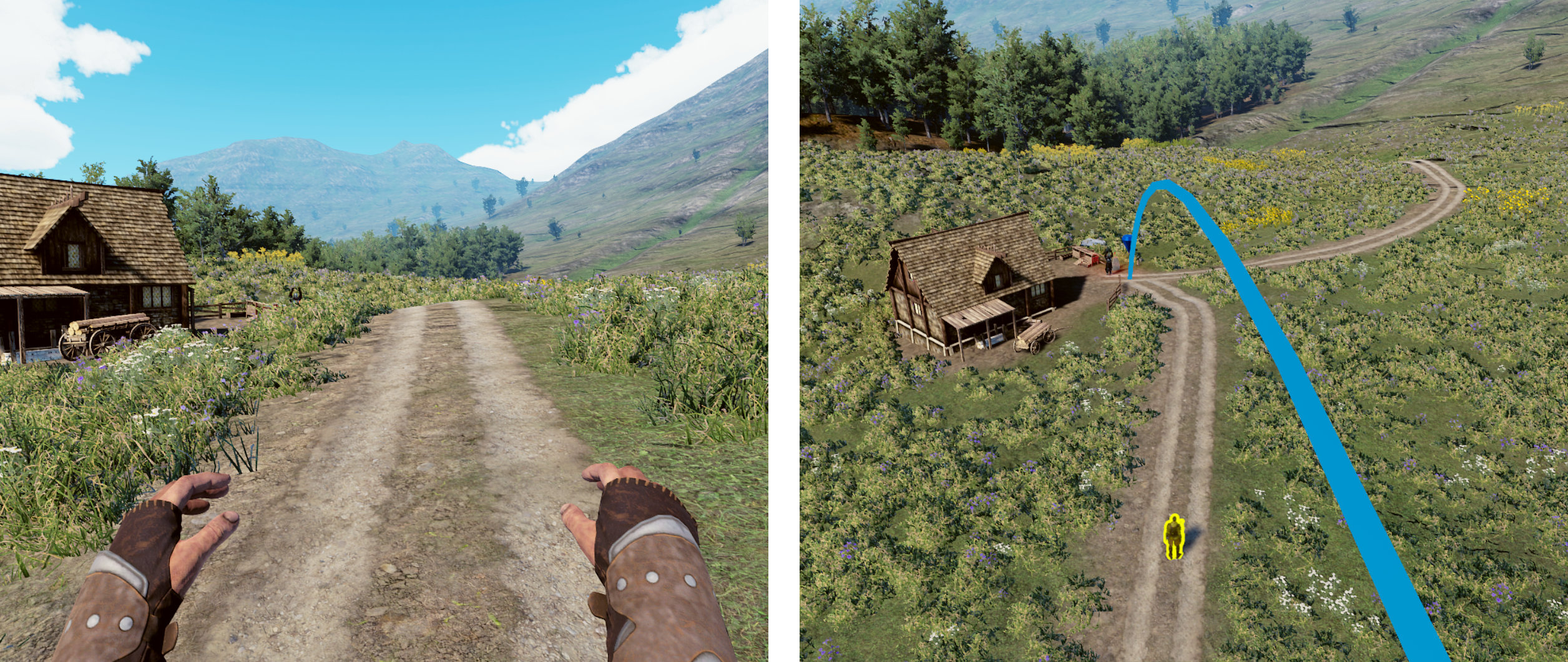}
\caption{\textit{Normal mode} (left) and \textit{travel mode} (right): In \textit{normal mode (NM)} (left), players use their virtual hands to interact with the environment. Upon switching to travel mode (TM) on demand, the players control their avatar from a third-person bird's eye perspective.}
\label{fig:outstanding}
\vskip -1em
\end{figure}

The main idea behind our locomotion technique is to switch between different perspectives on demand based on the current situation. In normal mode (NM), players perceive their surrounding from a first-person point of view. This perspective is used for short-range exploration by physical walking, detailed observation of local points-of-interest, and basic interactions such as picking up objects. In travel mode (TM), players leave their avatar behind and are scaled to a third-person bird's eye perspective. The virtual avatar is displayed at the players' feet symbolizing their original first-person position in the world. This view allows the disembodied players to observe the surrounding area from an elevated view and to command their avatar by setting navigation targets using raycast aiming (see Figure~\ref{fig:outstanding}). In TM, the players are completely decoupled from their avatar and are able to explore the world independently. Each perspective has benefits and drawbacks~\cite{gorisse2017first}: Third-person is excellent for environmental perception while first-person outperforms in interaction-intensive tasks. Through dynamic perspective switching, we combine the strengths of both views and achieve easy traveling and superior overview with local exploration and interaction on demand.\par
Our technique can be split into three components: normal mode, travel mode, and the transition between both states. We emphasize a proper design of such transitions, as they contribute to the players' spatial orientation and should not induce cybersickness. The naive approach would be an instantaneous switch between both perspectives~\cite{Habgood:2017:HLP:3130859.3131437}.\comm{ to avoid cybersickness through translating the virtual camera} However, this contradicts the primary goal of our approach to eliminate the immediate relocations known from arc-based teleport that could lead to disorientation. Instead, a fast automatic camera translation\comm{movement between both perspectives} is used\comm{ to convey a proper transition}, as earlier work~\cite{krekhov2018gullivr, Habgood:2017:HLP:3130859.3131437} showed that fast and brief movements do not induce cybersickness.\par

\begin{figure}
\centering
\includegraphics[width=1.0\columnwidth]{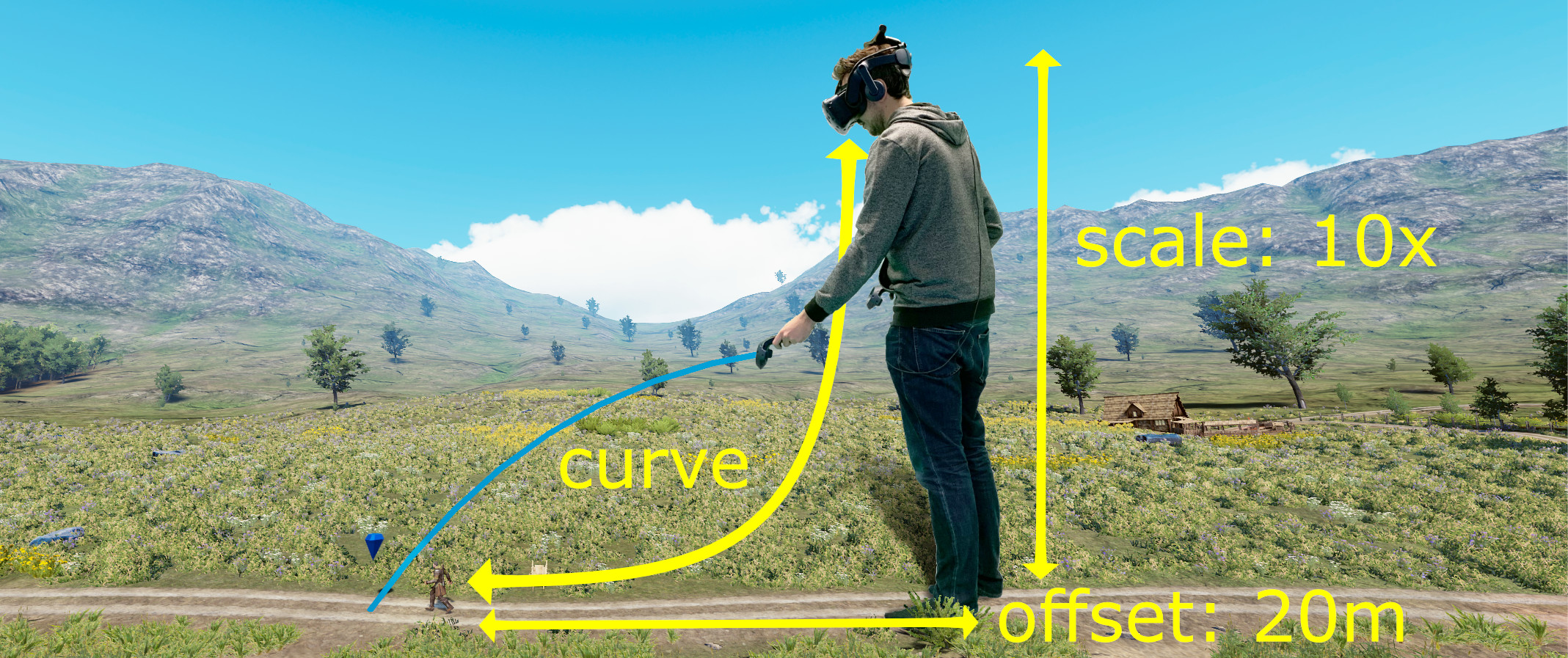}
\caption{The different transition parameters that were used to realize a continuous perspective switch and convey the feeling of embodying or disembodying the avatar.}~\label{fig:transitionParameters}
\end{figure}

Additionally, we extend the dolly-shot-like animation to improve the impression of embodying or disembodying the avatar. Early implementations kept the virtual position of the enlarged players so that they were located right above their avatar. However, this forced them to look straight down to see and command their character. A viewing angle of nearly 90\textdegree~is not only putting a strain on the human neck but also leads to a blurry vision through the headset as less pressure is applied to keep it tightly in place. Instead, we decided to add a translation backward to achieve a comfortable 45\textdegree~viewing angle after switching to TM. Furthermore, a curved animation between both states emphasized a \textit{horizontal disembodiment} followed by a steeper vertical growth (see Figure~\ref{fig:transitionParameters}). Early testers collectively approved this design decision as it felt more natural and matched the intended experience. As final extension, we implemented a previously requested feature: The possibility to speed up the avatar's movements to reduce the travel time.\par
\begin{figure*}[t!]
\centering
\includegraphics[width=2.1\columnwidth]{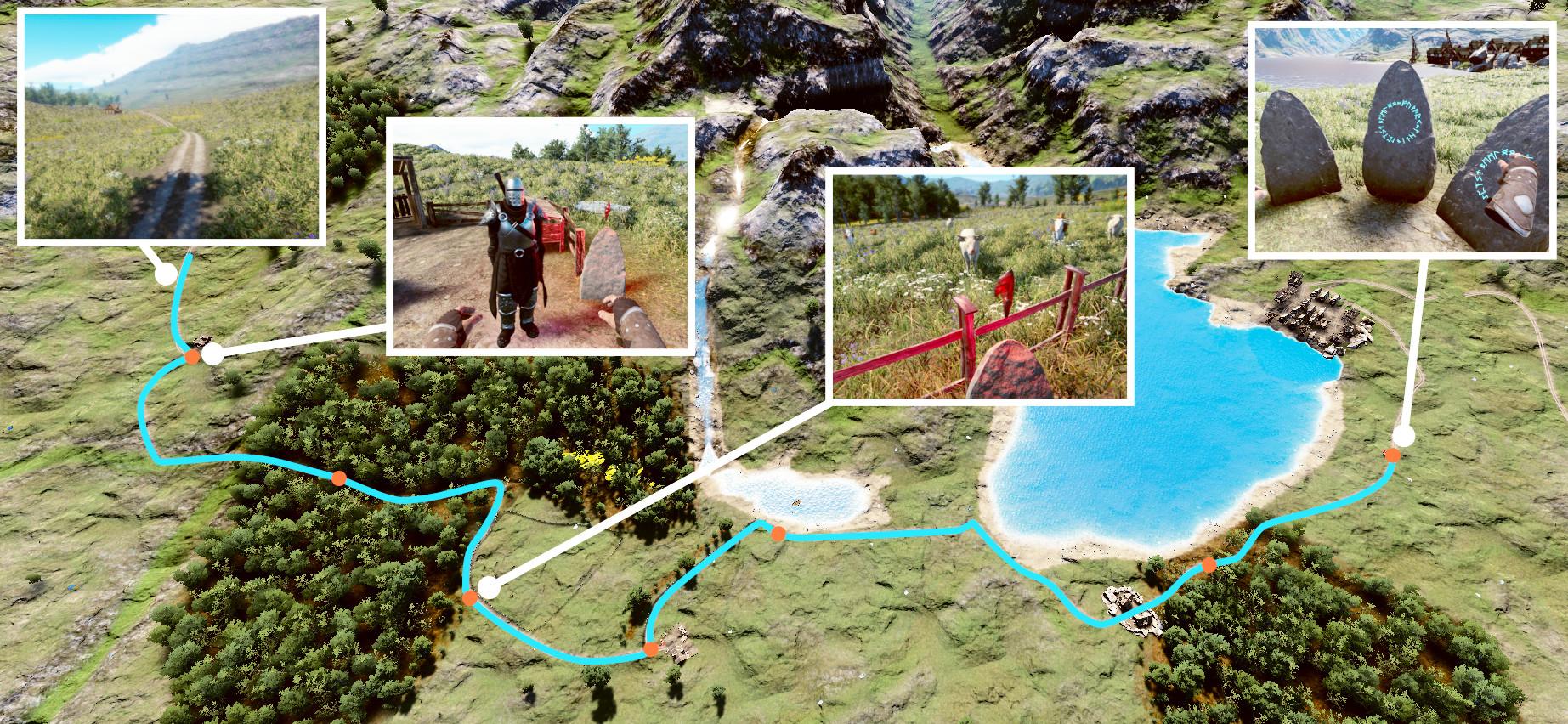}
\caption{Overview of the used scenario, including all important points of interest. From left to right: players follow on a path (Q1), meet a knight (Q2), destroy runes (Q3), and activate stones (Q4). The locations of all seven runes are marked as red dots.}
\label{fig:map}
\end{figure*}

\section{Evaluation}
We conducted a study to evaluate our proposed navigation technique. In this process, we designed a virtual environment with several square kilometers in size suited for testing large-scale locomotion and used it to compare the method against the most common and established alternative: the arc-based teleport. We were especially interested in how players would use the different mechanisms and how these would perform regarding performance, usability, orientation, and cybersickness. 

\subsection{Research Questions and Hypotheses}
Our main goal is to explore the difference between the two techniques. Since our approach is novel, our priority is to determine if players can complete all tasks regardless of the used locomotion approach. Furthermore, we are interested in whether there are any differences regarding the perceived presence, enjoyment, and competence. A particular focus is placed on the perspective-switching as it is a unique feature of our technique and has not yet been used for navigation approaches. Additionally, we hypothesize that our short and smooth animation curve, coupled with correctly altering the modeled eye distance, prevents the occurrence of cybersickness. Finally, we assume that the virtual scaling leading to an increased view height, provides significant advantages to spatial orientation and overview, as players can see much more and further. To summarize, our hypotheses and research questions are:

\begin{itemize}
  \setlength{\itemsep}{2pt}
  \setlength{\parskip}{0pt}
  \setlength{\parsep}{0pt}

\item H1: The perspective switching provides significant benefits to spatial orientation and overview.
\item H2: The proposed navigation through perspective-switching does not induce cybersickness.
\item RQ3: How do players perceive the perspective switch? Are they able to use this technique intuitively? 
\item RQ4: Are there any differences between both approaches in terms of perceived presence and enjoyment?
\item RQ5: How does our navigation technique perform regarding task completion and playtime against the arc-based teleportation?
\end{itemize}

\subsection{Scenario}
The game used to compare the different navigation techniques is realized using the Unity3D game engine~\cite{unity} and is set in a fictive, historical world including some fantasy aspects. The environment is a massive plateau-like scenery with multiple forests, lakes, and medieval towns. Everything is surrounded by a mountainous hinterland to achieve a restricted and enclosed play area. A long, wide, and twisted path connects all relevant locations and serves as the main point of orientation (see Figure~\ref{fig:map}). To use the different navigation techniques to a full extent, this path with adjacent points of interests is chosen to cover an extensive length of more than two kilometers. The main character, portrayed by the players, is a wandering mercenary and adventurer who just seeks the next unexpected incident. In the chosen scenario, the players are given the task to follow the path (\textbf{Q1}) that ultimately leads to the largest settlement on the map. However, after a short walk of about 150 meters, the players reach a forlorn farmhouse and a knight waiting for them. The knight asks the players to help him with a bigger quest: they have to destroy a red glowing rune floating above an obelisk next to the NPC (\textbf{Q2}). A harmless rabbit sits nearby and is under a spell from the rune. After the rune is picked up by the player, it dissolves, and the rabbit is freed. This first interaction allows the subjects to try out object manipulation through virtual grabbing. The knight asks the players to continue saving animals by following the path and destroying all other runes they encounter during their journey (\textbf{Q3}). In total, the game features seven distinct runes, each surrounded by different animals. This task is the central part of the quest and involves local object manipulation as well as large-scale navigation and long-distance orientation.\par
The last rune is located right in front of the large settlement serving as the final destination and floats in the middle of a circle of seven obelisks. After this last rune is destroyed, the knight reappears and asks the players to help him for the last time: They have to activate the seven obelisks surrounding them (\textbf{Q4}). To complete this last local task, the players have to approach each obelisk - usually by just stepping forward - and press their hand on the stone until a ring of engraved runes lights up. This task combines first-person interactions with short-range navigation. After activating all obelisks, the knight hands a sack of gold to the player and the game is completed.

\subsection{Procedure and Applied Measures}
We conducted a between-subject study splitting the group of participants randomly in two groups, each using either the arc-based teleport or our technique as navigation concept. This approach was chosen to avoid adverse sequence effect from repetition as one of the central use-cases for our technique is the exploration of unknown large-scale environments. The study was conducted in our VR lab using an HTC Vive Pro Wireless setup and took 50 minutes on average. We began by informing the participants about the overall procedure and administered a general questionnaire to assess gender, age, gaming behavior, and prior VR experience. Additionally, we assessed the ability to get immersed into games, books, or movies, by administering the Immersive Tendencies Questionnaire (ITQ)~\cite{Witmer.1998} consisting of the four subscales \textit{involvement}, \textit{focus}, \textit{games}, and \textit{emotions}. Finally, we introduced the subjects to the HTC Vive Pro and helped them to adjust the headset to their needs.\par
The game was preceded by a tutorial guiding the players through every critical aspect of VR games in general and our testbed scenario in detail. It included navigation-independent parts such as walking naturally in a room-scale environment or using the trigger buttons to grab and release objects. Additionally, the subjects were introduced to the particular navigation technique and had to use it to reach specified targets in the virtual world. After completing the tutorial, subjects were placed into the actual testbed game and given the main task. We logged all durations, interactions, and distances throughout the playthrough. Upon completion of the final quest, the players were asked to remove the head-mounted display.\par
As the final step in this study, we administered a series of questionnaires regarding the subjects' experiences. In order to assess the feeling of presence, we relied on two different questionnaires. The Igroup Presence Questionnaire (IPQ)~\cite{Schubert.1999b} focuses on general presence and contains one single item regarding the perceived general presence ("\textit{In the computer-generated world, I had a sense of 'being there'} "), as well as the three subdimensions \textit{spatial presence}, \textit{involvement}, and \textit{experienced realism}. For all of these items, participants had to rate statements on a 7-point Likert scale (coded 0 - 6). The Presence Questionnaire (PQ)~\cite{Witmer.1998,UQO.2004} focusing on interaction-related presence was used to determine the influence of the chosen navigation technique. It includes the items \textit{realism}, \textit{possibility to act}, \textit{quality of interface}, \textit{possibility to examine}, and \textit{self-evaluation of performance} (coded 0 - 6).
In order to measure the intuitiveness of the controls, we further administered the Player Experience of Need Satisfaction (PENS)~\cite{Ryan.2006} questionnaire with the subscales \textit{autonomy}, \textit{competence}, and \textit{intuitive controls}. Additionally, we included a single subscale of the Intrinsic Motivation Inventory (IMI)~\cite{ryan2000self}: \textit{interest/enjoyment} (coded 0 - 6).\par
Finally, we used the Simulator Sickness Questionnaire (SSQ)~\cite{Kennedy.1993} to examine whether the game in general or one of both navigation techniques might have caused cybersickness. The SSQ consists of the three subscales \textit{nausea}, \textit{oculomotor}, and \textit{disorientation} using a scale from 0 (none) to 3 (severe). The questionnaires were completed by several custom questions (coded 0 - 6) to gain essential insights into how the participants used the different navigation approaches (see Table~\ref{tab:Custom}). The study was finished by semi-structured interviews to allow all participants to share their experiences.

\section{Results}

\begin{table*}[t]
  \caption{Mean scores, standard deviations, and independent samples t-test values of the iGroup Presence Questionnaire (IPQ), the Presence Questionnaire (PQ), the Intrinsic Motivation Inventory (IMI) Questionnaire, and the Player Experience of Need Satisfaction (PENS) Questionnaire.}
  \label{tab:IPQ}
  \begin{tabular}{lccccl}
    \toprule
     & ~~~~~~~Outstanding ($N = 15$) & ~~~~~~~Teleportation ($N = 15$) \\
     \addlinespace 
      & ~~~~~~~M (SD)	& ~~~~~~~M (SD) & ~~~~~~~\textit{t} (28) & ~~~~~~~\textit{p} & \\
    \midrule
    PQ (scale: 0 - 6)\\
    \ \ \ \ \ \ Realism & ~~~~~~~4.12 (0.93) & ~~~~~~~4.05 (0.92) & ~~~~~~0.23 & ~~~~~~~.824&\\
    \ \ \ \ \ \ Possibility to Act & ~~~~~~~4.28 (0.65) & ~~~~~~~3.87 (0.62) & ~~~~~~~1.79 & ~~~~~~~.084&\\
    \ \ \ \ \ \ Interface Quality & ~~~~~~~4.53 (0.75) & ~~~~~~~4.69 (0.79) & ~~~~~~~-0.55 & ~~~~~~~.586&\\
    \ \ \ \ \ \ Possibility to Examine & ~~~~~~~4.31 (1.00) & ~~~~~~~4.71 (0.75) & ~~~~~~~-1.24 & ~~~~~~~.225&\\
    \ \ \ \ \ \ Performance & ~~~~~~~4.47 (1.23) & ~~~~~~~4.90 (0.74) & ~~~~~~~-1.17 & ~~~~~~~.252&\\
    \ \ \ \ \ \ Total & ~~~~~~~4.29 (0.62) & ~~~~~~~4.31 (0.55) & ~~~~~~~-0.08 & ~~~~~~~.935&\\
    IPQ (scale: 0 - 6)\\
    \ \ \ \ \ \ Spatial Presence & ~~~~~~~4.47 (0.60) & ~~~~~~~4.53 (0.89) & ~~~~~~~-0.24 & ~~~~~~~.812&\\
    \ \ \ \ \ \ Involvement & ~~~~~~~3.23 (1.31) & ~~~~~~~4.18 (1.31) & ~~~~~~~-1.99 & ~~~~~~~.057&\\
    \ \ \ \ \ \ Realism & ~~~~~~~2.57 (1.05) & ~~~~~~~2.60 (0.87) & ~~~~~~~-0.10 & ~~~~~~~.925&\\
    \ \ \ \ \ \ General & ~~~~~~~4.00 (1.25) & ~~~~~~~4.73 (1.10) & ~~~~~~~-1.70 & ~~~~~~~.100&\\
    IMI (scale: 0 - 6)\\
    \ \ \ \ \ \ Interest/Enjoyment & ~~~~~~~4.63 (0.74) & ~~~~~~~4.93 (0.96) & ~~~~~~~-0.96 & ~~~~~~~.345&\\
    PENS (scale: 0 - 6)\\
    \ \ \ \ \ \ Autonomy & ~~~~~~~2.87 (0.85) & ~~~~~~~3.49 (1.46) & ~~~~~~~-1.43 & ~~~~~~~.165&\\
    \ \ \ \ \ \ Competence & ~~~~~~~3.91 (1.33) & ~~~~~~~4.82 (1.24) & ~~~~~~~-0.47 & ~~~~~~~.640&\\
    \ \ \ \ \ \ Intuitive Controls & ~~~~~~~4.82 (0.60) & ~~~~~~~5.64 (0.48) & ~~~~~~~-4.14 & ~~~~~~~.000&**\\
    \bottomrule
     &&&& *\textit{p} <.05, ** \textit{p} <.01\\
\end{tabular}
\end{table*}
\begin{table}
  \caption{Mean scores and standard deviations of the Simulator Sickness Questionnaire (SSQ).}
  \label{tab:SSQ}
  \begin{tabular}{lcc}
    \toprule
    SSQ Dimension & ~~~~~~Outstanding~~~~~ & ~~~~~Teleport~~~~~\\
     & M (SD)	& M (SD) \\
    \midrule
    Nausea & 7.00 (8.43) & 14.63 (33.02) \\
    Oculomotor & 21.22 (16.26) & 20.72 (21.91)\\
    Disorientation & 25.06 (21.91) & 25.98 (37.89)\\
    Total & 19.95 (14.87) & 22.94 (32.63)\\
  \bottomrule
\end{tabular}
\end{table}

In total, 30 persons (9 female, 21 male) participated in our study with a mean age of $27.2$ ($SD=11.09$). Most participants reported playing digital games at least a few times a month. Even though the majority ($83\%$) had already used VR systems before, only $43\%$ of those reported using VR regularly. Therefore, the group of participants could be split nearly evenly into three categories: newcomers, occasionally users, and experienced VR gamers. All subjects were randomly split into two groups, one for each condition. These groups did not differ significantly regarding the distribution of VR experience, age, gender, or immersive tendencies ($t=0.44, p=.662$).\par
To answer our research questions and evaluate the hypotheses, we compared the results of all measures between the two conditions. To ensure the necessary requirements for parametric calculations, we tested for homogeneity of variances using Levene's and for normal distribution with Kolmogorov-Smirnov tests. If the requirements were not met, we replaced independent sample t-tests through Mann-Whitney U tests.\par

\subsection{Questionnaires}
In H2, we assumed that our technique avoids inducing additional cybersickness. The resulting weighted scale scores of the SSQ are depicted in Table~\ref{tab:SSQ}. It is to note that, concerning to reference values by Kennedy et al.~\cite{kennedy1993simulator}, all values are very low and indicate no problems with cybersickness in both conditions. Even though most dimensions show slightly better results for our approach in comparison to the control group, these differences are not significant (all $p>.393$). \par
In order to assess the research questions, we measured the perceived presence, competence, and enjoyment between the two study conditions. The resulting scores and independent sample t-tests are shown in Table~\ref{tab:IPQ}. The two presence questionnaires do not indicate any significant difference between our approach and the teleportation technique. However, two subscales stand out: players using the teleport tend to experience more involvement, while subjects in the other group report slightly more possibilities to act (both $p<.100$). Furthermore, most subscales show slightly better values for the teleport group. Similarly, the IMI questionnaire does not show any significance regarding the perceived enjoyment despite a slight tendency towards the established teleportation approach and very high values for both groups in general. Finally, the analysis of the PENS subscales reveals that both navigation approaches are perceived as reasonably intuitive. Nevertheless, the basic teleportation significantly outperforms the more complex perspective switching regarding the intuitiveness.

\begin{table*}[]
\caption{Mean scores and standard deviations of the custom questions (CQ) and independent samples t-test values of comparison.}
  \label{tab:Custom}
  \begin{tabular}{llccccl}
    \toprule
     && Outstanding & Teleportation &  &  &\\
     &Question Item & M (SD)	& M (SD) & \textit{t} (28) & Sig. \textit{p} & \\
    \midrule
    CQ1 & I would have preferred to move through the world \\ 
    &using another technique. & 3.27 (2.28) & 3.60 (2.06)  & 0.42 & .678 &\\\addlinespace
    CQ2& I think the navigation technique is intuitive. & 4.00 (1.46) & 4.60 (1.45) & 1.13 & .270 &\\\addlinespace 
    CQ3& I had problems with the supplied controls. & 2.20 (2.08) & 1.00 (1.36) & -1.87 & .074 &\\\addlinespace 
    CQ4& It was easy to reach the next destination. & 4.73 (1.49) & 4.87 (1.46) & 0.25 & .806 &\\\addlinespace 
    CQ5& I would have liked to have more variety during\\
    &the journey. & 3.73 (1.67) & 3.07 (1.87) & -1.03 & .312 &\\\addlinespace 
    CQ6& I felt very active while playing. & 4.07 (1.62) & 3.27 (2.09) & -1.17 & .251 &\\\addlinespace 
    CQ7& I could orient myself well in the game world. & 5.07 (0.88) & 3.67 (2.02) & -2.46 & .024 & *\\ \addlinespace 
    CQ8& After each relocation, I needed a moment\\
    &to orient myself. & 1.13 (1.73) & 3.80 (1.61) & 4.37 & .000 &**\\ \addlinespace 
    CQ9& From above, the world appeared to me as a \\&miniature or toy world. & 4.60 (1.60) & -- & --& -- &\\ \addlinespace
    CQ10& I liked the ability to experience the world\\& from above. & 4.73 (1.33) & -- & --& -- &\\ \addlinespace 
    CQ11& The perspective-switch confused me.& 0.67 (1.45) & -- & --& -- &\\ \addlinespace 
  \bottomrule
  &&&&*\textit{p} <.05, ** \textit{p} <.01
\end{tabular}
\end{table*}

\subsection{Custom Questions and Logging Data}
Apart from using standardized questionnaires to compare both groups, we assessed several custom questions to gain further insights into how players experience our technique. These questions covered intuitiveness, usability, orientation, and perspective-switching. The results shown in Table~\ref{tab:Custom} indicate a significant difference in spatial orientation: Participants using our presented approach reported being able to orient themselves far better and needing less time to recover from relocations. Additionally, the questions reveal several interesting trends between both groups.\par
We logged all relevant data that occurred during the play sessions (see Figure~\ref{fig:diagram}). In comparison to the teleport group, players using our technique played $66\%$ longer on average ($t(28)=6.00, p<.001$). This extended game time was mainly used to navigate in TM as the subjects played almost two-thirds of the complete game in third-person. One aim of our proposed approach was to enable longer and fewer aiming operations using raycast arcs. The result is a significant reduction in target aiming by $57\%$ in contrast to the control group ($t(16.63)=-4.72, p<.001$). In return, these saved interactions are replaced by the necessary switches between NM and TM. In sum, there is no significant difference in total user interaction count ($t(28)=-0.88, p=.388$). However, the type of aiming operations being used differs significantly: Players in the teleport-group mainly used medium-range teleportations exceeding the room scale-scope of two meters but mostly staying below $40$ to $50m$ at max. In contrast, our navigation technique enabled players to shift a huge amount of these medium-range arcs to a few long-distance travels. This advantage does not transfer to very near targets. In this case, the necessary aiming operations are nearly equal between both groups. Finally, we assessed the distance players walked in the real world while playing. While both study groups performed equally concerning the active walking in the first-person perspective ($t(28)=-0.67, p=.506$), i.e. approaching a rune, players using perspective switching walked $60\%$ more while commanding their avatar in travel mode ($t(28)=3.31, p=.003$). This closely reflects the results of CQ6.

\section{Discussion}

\textbf{RQ3:} \textit{How do players perceive the perspective switch? Are they able to use this technique intuitively? }\par
In general, most participants approved using different perspectives to experience the world from various views (CQ1). These were often described as \textit{"impersonating the avatar"(\textbf{P18})} versus switching to a \textit{"god-mode"(\textbf{P22})} and guiding a prot\'{e}g\'{e} through the world. The general concept of dynamic perspective switches was understood intuitively and did not confuse the subjects at all (CQ11). Instead, players appreciated the \textit{"balance between long-distance overview and local detailed exploration"(\textbf{P5})}.\par
One challenge our proposed technique introduces is the increased number of control mechanisms. In contrast to the teleport, our approach uses a second button to switch perspectives and an optional third control to speed up the travel process through running. These more complex interactions are one possible reason for the results from the PENS subscales and the custom questions. Subjects generally rated our approach as being less intuitive (CQ2) and reported more problems with the controls (CQ3). Especially new players stated that it \textit{"took some minutes not to confuse the different buttons anymore"(\textbf{P9})}. Nevertheless, a majority of players still reported being able to reach every destination easily (CQ4) after getting used to the technique.\par

\textbf{H1:} \textit{The perspective switching provides significant benefits to spatial orientation and overview.}\par
In TM, most participants perceived the environment as a toy world (CQ9). This perspective was generally appreciated (CQ10) as it placed the local surrounding \textit{"in greater context"(\textbf{P28})} and invited players to \textit{"explore the world"(\textbf{P2})}. This feedback fits the results of CQ7, revealing significant advantages to spatial orientation. Furthermore, the applied transition animation between TM and NM helped participants to preserve their cognitive map of the surrounding and reduced the necessary reorientation time after each switch (CQ8). These findings illustrate the most important advantage of our proposed technique: players can coordinate themselves better in a vast open world while avoiding to induce confusion through instant teleportation.

\begin{figure*}[t!]
\centering
\includegraphics[width=2.1\columnwidth]{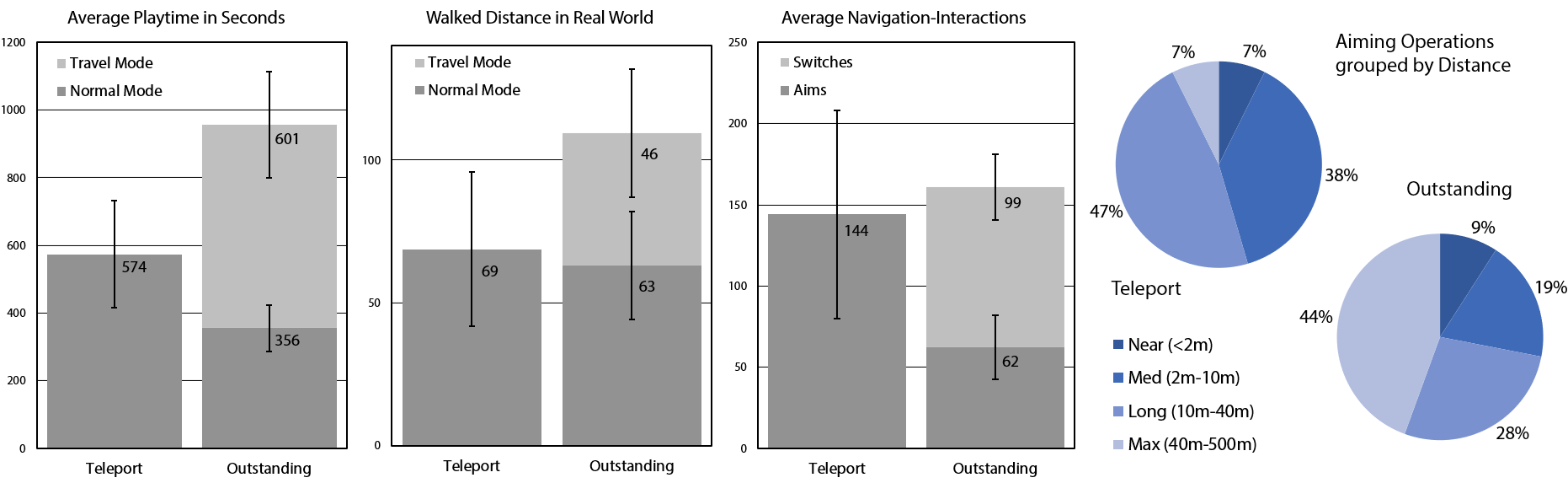}
\caption{Results from the data logged during the play sessions. From left to right: The difference in playtime for both study groups in seconds; The average distance (in meters) players walked in the real room; The difference in navigation-relation interaction count between both study groups; The distribution of aiming operations based on distance.}
\label{fig:diagram}
\end{figure*}
\begin{figure}
\centering
\includegraphics[width=1.0\columnwidth]{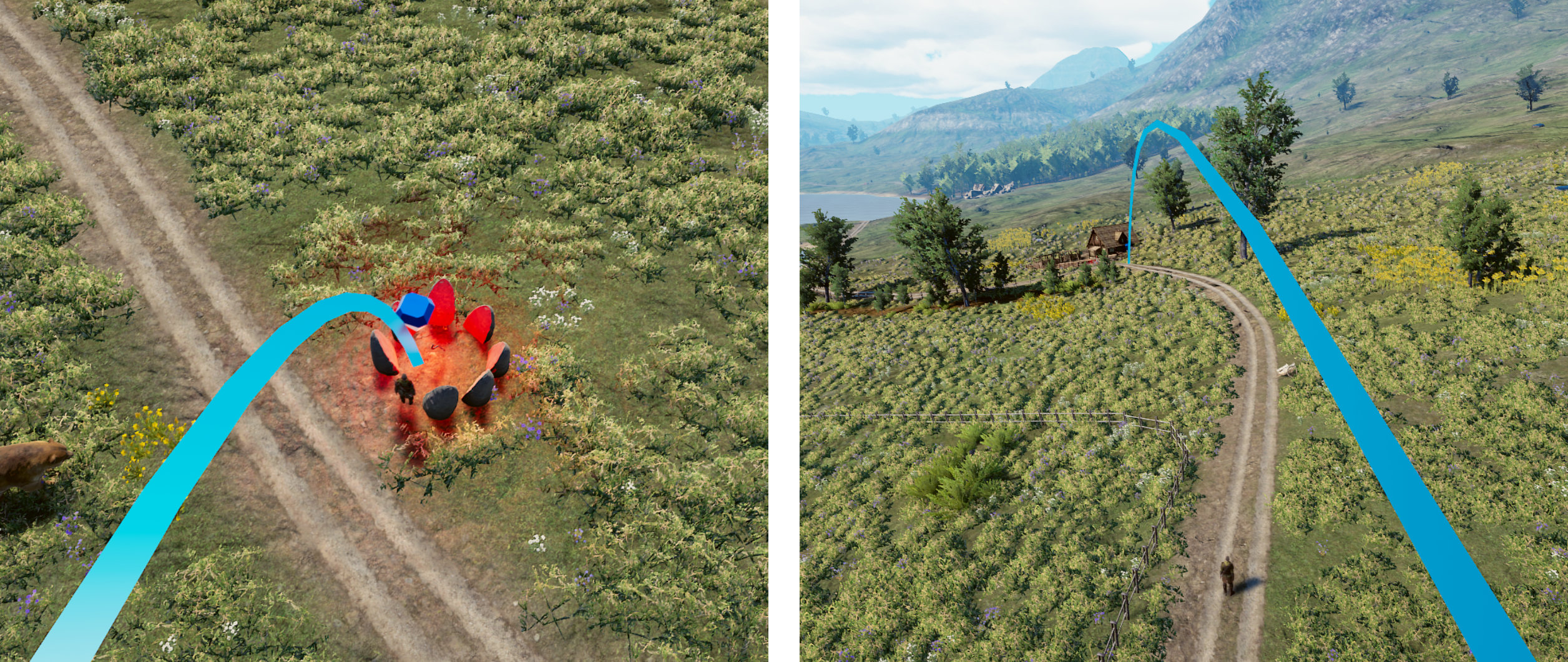}
\caption{Our proposed navigation technique performed differently depending on the distance: When traveling short distances (left), the overhead from switching perspectives outperformed the advantages. Longer travels (right) could be achieved through a single click.}
\label{fig:distances}
\end{figure}

\textbf{H2:} \textit{The proposed navigation through perspective-switching does not induce cybersickness.}\par
The results from the SSQ indicate that our technique does not negatively affect players in terms of cybersickness. Even though our approach includes an automated virtual movement that contradicts the signals from the vestibular system, this does not cause any symptoms. This positive finding is in line with earlier work~\cite{krekhov2018gullivr, Habgood:2017:HLP:3130859.3131437}, emphasizing the importance of short and fast movements. Additionally, we eliminate any side effects arising from the altered perception in TM by scaling the virtual eye distance accordingly to the body size.

\textbf{RQ4:} \textit{ Are there any differences between both approaches in terms of perceived presence and enjoyment?}\par
In general, the two navigation techniques did not differ significantly concerning presence. However, the teleport was mostly rated slightly higher than our proposed approach, which is especially true for the IPQ subscale \textit{involvement}. This result fits the general feedback: Players did not have the feeling of controlling their avatar from a third-person perspective but felt like \textit{"disembodied beings guarding a traveler on his path"(\textbf{P24})}. The results from PQ and IPQ and the verbal feedback show that the participants were less involved while using the TM to travel through the world.\par
Even the best interaction technique will never be adopted in games if it does not provide a compelling and fun experience. On first sight, the \textit{interest/enjoyment} subscale of the IMI questionnaires reveals slightly lower results for the perspective switching approach in comparison to the teleport. However, when comparing the techniques more closely, two aspects become clear: First, the general scores for both groups are very high, and subjects in the lower rated Outstanding-group played roughly $66\%$ longer. Together, these findings illustrate that the proposed approach preserves nearly equal levels of enjoyment despite a significantly extended playtime.

\textbf{RQ5:} \textit{How does our navigation technique perform regarding task completion and playtime against the arc-based teleportation?}\par
All participants were able to complete the presented tasks and ultimately finish the game. However, subjects using our proposed navigation technique needed significantly more time for the same quests. Most of this overhead is due to the natural avatar walking speed: Even without any pause or switch to NM, the virtual avatar would need nine minutes of continuous walking or four minutes of running to reach the final destination. This difference in playtime does not necessarily imply a worse performance, as the game did not issue a time-relevant task. Instead, players were free to travel the world at their own speed. However, most subjects still requested either faster travels or more varieties during the journey: The presented scenario did not include enough point of interest for 15-minute gameplay. We propose to enrich the virtual world through interesting spots that make traveling a compelling experience.\par
One of the minor motivations to develop this navigation alternative was to reduce the large number of aiming operations that are necessary when using the teleport to traverse long distances. This goal was only partly fulfilled: For small distances, e.g., reaching a rune nearby, the perspective-switching technique did not lower the necessary aiming operations (see Figure~\ref{fig:diagram}, right) but imposed a minor overhead through the necessity to switch between NM and TM. We, therefore, conclude that our approach should be accompanied by an alternative fallback technique such as the teleport for very short travels. Nevertheless, the analysis reveals a major reduction of raycast-arcs in the medium-range between $2 and 40 meters$ that were replaced by fewer very long travel operations (see Figure~\ref{fig:distances}). This finding underlines our initial assumption: the elevated point of view enables players to aim further and travel large distance with the ease of one command.\par

\section{Design Implications}
The proposed navigation technique introduces dynamic perspective switches and extends the collection of the existing locomotion approaches for VR. This section presents a set of design implications regarding the use of our approach for VR games and applications.

\subsubsection{Combination with other Navigation Techniques}
\comm{One central goal of our proposed technique was to make travels easier and reduce the necessary effort of multiple teleportations. While this goal was reached for long distances, the perspective switching does not provide many benefits for local areas. Instead, it adds an additional overhead through the necessary switches between NM and TM. Therefore, we propose to accompany our approach through an additional navigation technique for short ranges, e.g., the teleport.}
One central goal of our proposed technique was to make travels easier and reduce the necessary effort of multiple teleportations. The study undermines the benefits of perspective switching for long distances. However, the necessary interactions add additional overhead for local areas. Therefore, we propose to accompany our approach through an additional navigation technique for short ranges, e.g., the teleport.

\subsubsection{Choosing the Parameters}
In our early design phase, we administered multiple variants of the parameters, such as TM scaling size, transformation curve, and horizontal offset. Even though these values proved to be optimal for our setup, other use cases could potentially make adjustments necessary.
\begin{itemize}
  \setlength{\itemsep}{2pt}
  \setlength{\parskip}{0pt}
  \setlength{\parsep}{0pt}
\item Size: In our scenario, we used a predefined scale of $10x$ to preserve most details while still benefiting to a general overview and better aiming. Participants generally appreciated this balance, even though it introduced additional challenges, i.e., avatar visibility in dense forests. A possible solution would be to add user-defined sizes, though these introduce additional degrees of freedom and generally perform worse than predefined values~\cite{kopper2006design}.
\item Transformation: The transformation between NM and TM should be chosen fast enough to avoid inducing cybersickness and slow enough to convey the feeling of embodiment. In our early design phases, values around half a second were rated best. Additionally, a curved transformation (see Figure~\ref{fig:transitionParameters}) was preferred over a linear movement and growth, as it felt smoother and more natural. Another critical aspect to a successful perspective switch is the correctly modeled eye distance. Misalignments in the so-called stereo base easily lead to strong symptoms of cybersickness.
\item Horizontal offset: Initially, we used a linear vertical growth and placed the avatar right to the players' feet. However, looking straight down induces severe strain on the users' necks and often leads to blurry vision as the HMDs are not entirely fixated. Instead, we propose an additional horizontal offset backward (see Figure~\ref{fig:transitionParameters}) so that the players can see their avatar at a comfortable 45\textdegree angle.
\end{itemize}

\subsubsection{Closed Areas and Vertical Level Design}
Naturally, our proposed technique does not work very well with closed ceilings as players are scaled to a bird's eye perspective and would clip through the roof while switching to TM. However, this problem is solvable for environments that are not restricted to indoor areas, such as caves. In the popular case of a house in an open-world scenario, it would be easy to hide the roof or parts of the walls while the players are in TM. This tweak would enable them to control their avatar in the house even for multi-story buildings while standing outside of the house and 'looking' through the walls.

\subsubsection{Catching Up}
Many participants suggested an additional control mechanism to catch up on the avatar. Usually, they sent him to a far-away target and had to switch between TM and NM regularly to accompany him during the journey. Since this provided an unnecessary overhead, they wished to be able to skip ahead and close the gap between the player and the avatar without switching perspectives. However, overuse of this artificial transition process could easily evoke cybersickness and should be used cautiously and sparsely.\par

\subsubsection{Avatar Visibility}
In areas like dense forests, players sometimes lost sight of their avatar due to occlusions by obstacles, such as trees or rocks. This made it hard to set proper navigation targets and follow the requested path. This drawback could be solved by culling or fading occluding objects or highlighting the avatar with an outline. Another solution would be to increase the virtual scaling of the travel mode. %

\subsubsection{Higher Point of View}
While providing superior orientation and overview, the higher point of view in TM raises additional design challenges as well. The player could easily get spoiled from seeing too far ahead. This issue could be tackled by various possible solutions, e.g., placing natural obstacles such as mountains or using artificial techniques like the fog of war. 

\subsubsection{Providing Variety during Travel}
One of the biggest requests to make our approach more compelling was the variety during longer journeys. Players had to wait for their avatar to reach his target and had quickly seen everything of their surrounding. Usually, digital games try to keep the players engaged by introducing new events regularly. In our case, a walk of two to three minutes is likely too long to preserve or increase the perceived enjoyment. Therefore, we propose to add additional incidents, either first-person encounters that force the players to switch back to NM or special actions that can be completed while waiting for the avatar, e.g., removing road barriers that stop the avatar from reaching his target. Such interactions could provide novel game mechanics and possibilities. 

\subsubsection{Novel Player Experiences}
Even though the participants intuitively understood the concept of switching between the two perspectives, they commonly reported not having the feeling of controlling their own avatar. Instead, it was perceived more as a protector-prot\'{e}g\'{e}-relationship: The players controlled an avatar and could possess him whenever necessary. This experience is basically a flipped understanding of the underlying perspective-switch and could be used as a novel game mechanics, e.g., by controlling multiple characters at once. Additionally, the participants emphasized the importance of creating coherence between both perspectives to achieve a plausible transition.\par
Another interesting side effect of our approach is the impression of a miniaturized world. The proportional increase in modeled eye distance leads to perceiving the surrounding as a toy world, which allows equipping players with respective "godlike" abilities, such as increased strength. One possible application would be special TM interactions: For instance, certain obstacles, e.g., logs or boulders, could be too heavy for the first-person character and would need to be lifted in TM.

\subsubsection{Possible Applications}
We asked the participants to name potential games that would profit from our navigation approach. The most commonly named genres were adventures, role-playing games, and strategy games. We consider that especially slow titles relying on huge open worlds are suited best. In general, the perspective switching is too slow for fast-paced gameplay, as in first-person shooters. Apart from VR gaming, our proposed technique could provide essential benefits for scenarios requiring spatial orientation. Examples for such applications are large VR exhibitions or environmental visualizations.

\section{Conclusion and Future Work}
The variety of established navigation techniques for virtual environments is immense. Nevertheless, there is currently no perfect approach for compelling and continuous travel over large distances. Our presented technique is a novel alternative based on dynamic perspective switching. Players can interact with the world on a local scale and switch to a third-person travel mode on demand. In contrast to the prominent teleport technique, our approach increases spatial orientation in large worlds while avoiding cybersickness and preserving high levels of presence, competence, and enjoyment. Our experiments showed that players generally liked the idea of the dynamic switching between different perspectives and that they were able to use the technique without major problems. Additionally, we summarized the key insights from the various measures and verbal feedback into a set of comprehensive design guidelines.\par
In our future research, we will focus on improvements and applications of our technique. A special focus will be placed on altering the approach for additional use cases, such as indoor traveling. Another interesting research question that could provide essential insights towards playing multiple characters at once, is the relationship between the players and their avatars. Furthermore, we suggest to combining Outstanding with alternative techniques for close-range navigation. Finally, we propose to investigate the potential use cases of our technique for different game genres and VR applications in general. 

\balance{}

\bibliographystyle{SIGCHI-Reference-Format}
\bibliography{Literatur}

\end{document}